\documentclass[aps,prb]{revtex4}
\usepackage{graphicx}

\begin{document}
\pagestyle{empty}
\title{Near-field radiative heat transfer  between closely spaced
graphene and amorphous SiO$_2$}

\author{A.I.Volokitin$^{1,2}$\footnote{Corresponding author.
\textit{E-mail address}:alevolokitin@yandex.ru}    and B.N.J.Persson$^1$}
 \affiliation{$^1$Peter Gr\"unber Institut,
Forschungszentrum J\"ulich, D-52425, Germany} \affiliation{
$^2$Samara State Technical University, 443100 Samara, Russia}

\begin{abstract}
We study the near-field radiative energy transfer  between graphene
and an amorphous SiO$_2$ substrate. In comparison with the
existing theories of near-field radiative heat transfer our theory
takes into account that the free carriers in graphene are moving
relative to the substrate with a drift velocity $v$. In this case the heat flux is
determined by both thermal and quantum fluctuations. We find that quantum fluctuations give
an important contribution to the radiative energy transfer for low temperatures and high electric field (large drift velocities).
For nonsuspended graphene  the near-field radiative energy transfer gives a significant contribution to
the heat transfer, in addition to the contribution from phononic coupling.     For
suspended graphene (large separation) the
corresponding radiative energy transfer coefficient  at nanoscale gap is $\sim$ 3
orders of magnitude larger than radiative heat transfer coefficient of the blackbody radiation
limit.
\end{abstract}

\maketitle


\vskip 5mm

 Transfer of energy between two surfaces separated by vacuum gap is a topic that has fascinated several  generations of researches. If both surfaces
 are at rest then at large separation $d\gg
\lambda_T=k_BT/\hbar$ the radiative heat transfer is determined by
the Stefan-Boltzman law, according to which the thermal heat transfer coefficient
 $\alpha=4\sigma T^3$. In this limiting case the heat
transfer between two bodies is determined by the propagating
electromagnetic waves radiated by the bodies, and does not depend
on the separation $d$. At $T=300$K this law predicts the (very
small)  heat transfer coefficient, $\alpha \approx 6$
Wm$^{-2}$K$^{-1}$. However, as was first predicted theoretically
by Polder and Van Hove \cite{Polder1971} in the framework of
stochastic electrodynamics introduced by Rytov
\cite{Rytov53,Rytov67,Rytov89}, and recently confirmed
experimentally \cite{Chen2009a,Greffet2009}, at short separation
$d\ll \lambda_T$, the heat transfer may increase by many orders of
magnitude due to the evanescent electromagnetic waves; this is
often referred to as photon tunneling. Particularly strong
enhancement occurs if the surfaces of the bodies can support
localized surface modes such as surface plasmon-polaritons,
surface phonon-polaritons, or adsorbate vibrational modes
\cite{Greffet6,VolokitinRMP07}.

The theory of the radiative heat transfer developed in Ref.
\cite{Polder1971} is only valid for bodies at rest. A more general
theory of the radiative energy transfer between moving bodies, with
arbitrary relative velocities, was developed by us in Ref.
\cite{Volokitin2008b}. According to this theory there is transfer
of energy between moving bodies even at zero temperature
difference, and the heat is generated by the relative
motion of quantum and thermal fluctuations. It appear in its most elementary form when the surfaces are at zero
Kelvin and the heat is generated  by the relative movement of quantum fluctuations. In this Communication this theory is
applied to calculate the radiative energy transfer between carriers
(moving with the drift velocity $v$) in graphene and the
substrate.

Graphene, the recently isolated single-layer carbon sheet, consist
of carbon atoms closely packed in a flat two-dimensional crystal
lattice. The unique electronic and mechanical properties of
graphene\cite{Geim2004, Geim2005} is being actively explored both
theoretically and experimentally because of its importance for
fundamental physics, and for possible technological applications
\cite{Geim2007}. In particular, a great deal of attention has been
devoted to the applications of graphene for electronics and
sensors \cite{Geim2004,Geim2007}.

For nonsuspended graphene direct phononic coupling also
contributes to heat transfer \cite{Pop2011,Persson2010a,Persson2010b,Persson2010c}. Graphene interact very weakly with
most substrates mainly via van der Waals forces. According to
theoretical calculations\cite{Persson2010a,Persson2010b,Persson2010c} the heat transfer coefficient due to the direct phononic coupling for the interface
between graphene and a perfectly smooth (amorphous) SiO$_2$
substrate is $\alpha_{\rm ph} \approx 3\times
10^8$Wm$^{-2}$K$^{-1}$, and according to experiment\cite{Chen2009}
(at room temperature)  the heat transfer coefficient ranges from
8$\times$10$^7$ to 1.7$\times$10$^8$Wm$^{-2}$K$^{-1}$ (however,
these values are probably influenced by the substrate surface
roughness).

In this Communication we investigate heat generation and dissipation due
to  friction produced by the interaction between   moving (drift velocity $v$) charge
carriers in graphene and the optical phonons in nearby
amorphous SiO$_2$, and the acoustic phonons in graphene. Friction produces work and thermal heating of the graphene which  results in
near-field radiative energy transfer and phononic heat transfer between the graphene and SiO$_2$.
 A self-consistent theory that describes these phenomena was  formulated by us in
Ref. \cite{Volokitin2011PRL} and it
allows us to predict experimentally measurable effects.  In
comparison with the existing microscopic theories of energy transfer and transport in
graphene \cite{Perebeinos2009,Perebeinos2010}  our theory is
macroscopic. The electromagnetic interaction between graphene and
a substrate is described by the dielectric functions of the
materials which can be accurately determined from theory and experiment.

 Consider graphene and a substrate, with
flat parallel surfaces at separation $d\ll \lambda_T=c\hbar/k_BT$.
Assume that the free charge carriers in graphene move  with the
velocity $v\ll c$  ($c$ is the light velocity) relative to the
substrate. According to Ref.
\cite{Volokitin2008b} the frictional
stress $F_x$ acting on the charge carriers in graphene  and the
radiative energy flux $S_z$ across the  substrate surface, both
mediated by a fluctuating electromagnetic field, are determined by

\[
F_x =\frac \hbar {\pi ^3}\int_{0 }^\infty dq_y\int_0^\infty
dq_xq_xe^{-2qd}\Bigg \{ \int_0^\infty d\omega \Bigg(
\frac{\mathrm{Im}R_{d}(\omega)\mathrm{Im}R_{g}(\omega^+) }{\mid
1-e^{-2 q d}R_{d}(\omega)R_{g}(\omega^+)\mid ^2}\times
\]
\[
 [n_d(\omega )-n_g(\omega^+)]+\frac{\mathrm{Im}R_{d}(\omega^+)\mathrm{Im}R_{g}(\omega^) }{\mid
1-e^{-2 q d}R_{d}(\omega^+)R_{g}(\omega)\mid ^2}[n_g(\omega
)-n_d(\omega^+)]\Bigg )+
\]
\begin{equation}
 \int_0^{q_xv}d\omega \frac{\mathrm{Im}R_{d}(\omega)\mathrm{Im}
R_{g}(\omega^-)} {\mid 1-e^{-2qd}R_{d}(\omega)R_{g}(\omega^-)\mid
^2} [n_g(\omega^-)-n_d(\omega)] \Bigg \}, \label{force}
\end{equation}
\[
S_z =\frac \hbar {\pi ^3}\int_{0 }^\infty dq_y\int_0^\infty
dq_xe^{-2qd}\Bigg \{ \int_0^\infty d\omega \Bigg(- \frac{\omega
\mathrm{Im}R_{d}(\omega)\mathrm{Im}R_{g}(\omega^+) }{\mid 1-e^{-2
q d}R_{d}(\omega)R_{g}(\omega^+)\mid ^2}\times
\]
\[
 [n_d(\omega )-n_g(\omega^+)]+\frac{\omega^+\mathrm{Im}R_{d}(\omega^+)\mathrm{Im}R_{g}(\omega^) }{\mid
1-e^{-2 q d}R_{d}(\omega^+)R_{g}(\omega)\mid ^2}[n_g(\omega
)-n_d(\omega^+)]\Bigg )+
\]
\begin{equation}
 \int_0^{q_xv}d\omega \frac{\omega \mathrm{Im}R_{d}(\omega)\mathrm{Im}
R_{g}(\omega^-)} {\mid 1-e^{-2qd}R_{d}(\omega)R_{g}(\omega^-)\mid
^2} [n_g(\omega^-)-n_d(\omega)] \Bigg \}, \label{heat}
\end{equation}
where  $n_i(\omega )=[\exp (\hbar \omega /k_BT_i)-1]^{-1}$
($i=g,d$), $T_{g(d)}$ is the temperature of graphene (substrate),
 $R_{i}$  is the reflection amplitude for
surface $i$ for $p$ -polarized electromagnetic waves, and
$\omega^{\pm}=\omega \pm q_xv$.  The reflection amplitude for
graphene (substrate)  for $q\gg \sqrt{|\epsilon_d(\omega)|}\omega/c$ is determined by \cite{Volokitin2001b}
\begin{equation}
R_{g(d)}=\frac{\epsilon _{g(d)}-1}{\epsilon _{g(d)}+1},
 \label{refcoef}
\end{equation}
where $\epsilon _{g(d)}$ is the dielectric function for graphene
(substrate). In the study below we used the dielectric function of graphene, which was
calculated recently within the random-phase approximation (RPA)
\cite{Wunsch2006,Hwang2007}. The different pieces of this dielectric function on  real axis in the complex plane can be obtained from a complex-
valued function which is  analytical  in the upper half-space
of the complex $\omega$-plane \cite{Volokitin2011PRL}. The dielectric function of amorphous SiO$_2$ can be described
using an oscillator model\cite{Chen2007}.

According to Eqs. (\ref{force}) and (\ref{heat})  in the case when free carriers are moving relative to the substrate both thermal
and quantum fluctuations give contributions to the frictional
stress and the radiative energy transfer . This situation
is  different from that considered in \cite{Persson2010a,Persson2010b} where it was assumed that the free carries in graphene had vanishing drift velocity. The contribution of
the quantum fluctuations to the frictional stress was investigated by us in  Ref. \cite{Volokitin2011PRL}. According to Eq. (\ref{heat}) the contribution to
 the near-field energy transfer from
quantum fluctuations is determined by

\begin{equation}
S_z^{quant}=S_z(T_d=T_g=0) =-\frac\hbar {\pi ^3}\int_{0 }^\infty
dq_y\int_0^\infty dq_x\int_0^{q_xv}d\omega \omega e^{-2qd}
\frac{\mathrm{Im}R_{d}(\omega)\mathrm{Im} R_{g}(\omega^-)} {\mid
1-e^{-2qd}R_{d}(\omega)R_{g}(\omega^-)\mid ^2}
\end{equation}

The  steady state temperature can be obtained from the condition that
the power generated by friction must be equal to the energy transfer
across the substrate surface
\begin{equation}
F_{t}(T_d,T_g)v=S_z(T_d,T_g)+\alpha_{ph}(T_g-T_d), \label{eqtemp}
\end{equation}
where $F_{t}$ is the total friction force which is  the  sum of  the extrinsic friction force $F_x$ due to interaction with optical
phonons in SiO$_2$, and the intrinsic friction force due interaction with acoustic and optical phonons in graphene.  The friction force
due to interaction with acoustic phonons at low velocities is determined by: $F_{ac}=ne\mu^{-1}v$, where $\mu$ is low-field mobility due to scattering
of the carriers against
 the acoustic phonons of graphene. At room temperature $\mu\approx$20 m$^{-2}$V$^{-1}$s$^{-1}$ \,\,\cite{ChenJH2008}. At other temperatures
 the mobility can be obtained taking into account that $\mu^{-1}$ depends approximately linearly on $T_g$. The friction force acting on the charge
 carriers in graphene for
high electric field (large velocities) is determined by the interaction with the
optical phonons of the graphene, and with the optical phonons of
the substrate. The frequency of optical phonons in graphene is a
factor 4 larger than for the optical phonon in SiO$_2$. Thus, one
can expect that for graphene on SiO$_2$ the high-field friction force will be determined by excitations of optical
phonons in SiO$_2$. The second term on the right side of  Eq. (\ref{eqtemp}) takes into account the
heat transfer through direct phononic coupling; $\alpha_{ph}$
is the heat transfer coefficient due to phononic coupling. Due to weakness of the van der Waals interaction between graphene and
substrate the application of the theory outlined above is justified for both suspended and nonsuspended graphene.

\begin{figure}
\includegraphics[width=0.80\textwidth]{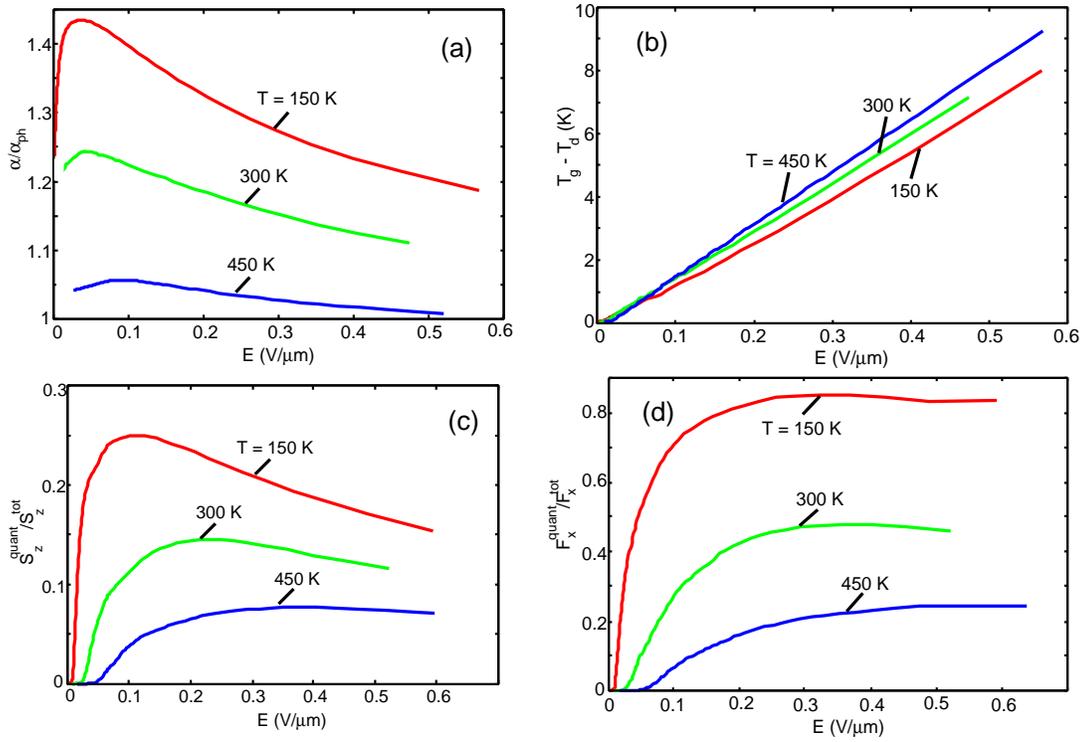}
\caption{\label{Fig.1.} Radiative energy transfer between graphene   and
SiO$_2$ for $n=10^{16}$m$^{-2}$, $d=0.35$ nm and $\alpha_{ph}=1.0\times
10^8$Wm$^{-2}$K$^{-1}$. (a) The dependence of the ratio between the total energy
transfer coefficient and the phononic heat transfer coefficient, on electric
field.  (b) Dependence of the temperature difference
between graphene and substrate on the electric field.
 (c) Dependence of the ratio between the heat flux only due to quantum fluctuations $S_z^{quant}$ and the
total  energy flux, on the electric field. (d) Dependence of the ratio between the friction force  only due to quantum fluctuations $F_x^{quant}$ and the
total friction force, on the electric field.}
\end{figure}

As discussed above, for graphene on SiO$_2$ the
excess heat generated by the current is transferred to the substrate
through the near-field radiative heat transfer,
and via the direct phononic coupling (for which the heat transfer coefficient $\alpha \approx 10^8$Wm$^{-2}$K$^{-1}$).  At
small temperature difference ($\Delta T=T_g - T_d\ll T_d$), from
Eq. (\ref{eqtemp}) we get
\begin{equation}
\Delta T=
\frac{F_{x0}v-S_{z0}}{\alpha_{ph}+S_{z0}^{\prime}-F_{x0}^{\prime}v}
\label{eqtemp1}
\end{equation}
where $F_{t0}=F_t(T_d,T_g=T_d)$, $S_{z0}=S_z(T_d,T_g=T_d)$,
\[
F_{t0}^{\prime}=\frac{dF_t(T_d,T_g)}{dT_g}\Big|_{T_g=T_d}, \,\,
S_{z0}^{\prime}=\frac{dS_z(T_d,T_g)}{dT_g}\Big|_{T_g=T_d}
\]
We note that, in contrast to the heat transfer between bodies at rest, for moving bodies the energy flux $S_z(T_d,T_g)$ is not equal to zero even for the case when there is no temperature difference between the bodies. The energy transfer coefficient is given by
\begin{equation}
\alpha =\frac {S_z(T_d,T_g)+\alpha_{ph}\Delta T}{\Delta T}
\approx
\frac{(\alpha_{ph}+S_{z0}^{\prime})F_{t0}v-S_{z0}F_{t0}^{\prime}v}{F_{t0}v-S_{z0}}
\label{alpha}
\end{equation}
For small velocities $F_{t0}\sim v$ and $S_{z0}\sim v^2$. Thus from Eq. (\ref{alpha}) it follows that
in the limit $v\rightarrow 0$ the energy transfer coefficient between moving bodies is not reduced to the heat transfer
coefficient between bodies at rest which is determined by $\alpha_{th}=\alpha_{ph}+S_{z0}^{\prime}$. This effect is due to the term $S_{z0}$ in the total
energy flux which exists only between moving bodies. The energy transfer coefficient can be strongly enhanced in comparison
to the heat transfer coefficient when $F_{t0}v\approx S_{z0}$.
Fig. \ref{Fig.1.}a shows the ratio of the energy transfer coefficient to the phononic heat transfer coefficient for $d=0.35$ nm and $n=10^{16}$ m$^{-2}$.
 For low and intermediate field this ratio is larger than unity what means that in this region the near-fields
 radiative energy transfer gives additional significant contribution to the heat transfer due to direct phononic coupling.
For nonsuspended graphene on SiO$_2$ the energy and heat transfer are very effective and the temperature difference does not rise high, even
for such high electric field that  saturation in $I-E$ characteristic starts \cite{Freitag2009} (see Fig. \ref{Fig.1.}b). The radiative
heat transfer between bodies at rest is determined only by thermal fluctuations, in contrast to the radiative energy transfer between
moving bodies which is determined by both thermal and quantum fluctuations.  Fig. \ref{Fig.1.}c shows that quantum fluctuations can give
significant contribution to the total energy transfer for low temperatures and large electric field (high drift velocity). Similarly,
in the (elecric current) saturation region quantum fluctuations give significant contribution to the total friction force which is determined, as discussed
above, by the sum of the extrinsic and intrinsic friction forces (see  Fig. \ref{Fig.1.}d) . The extrinsic friction force has contributions
from both thermal and quantum  fluctuations. The friction force due to quantum fluctuations is usually named as quantum friction which
was discussed by us recently in Ref. \cite{Volokitin2011PRL}.

\begin{figure}
\includegraphics[width=0.80\textwidth]{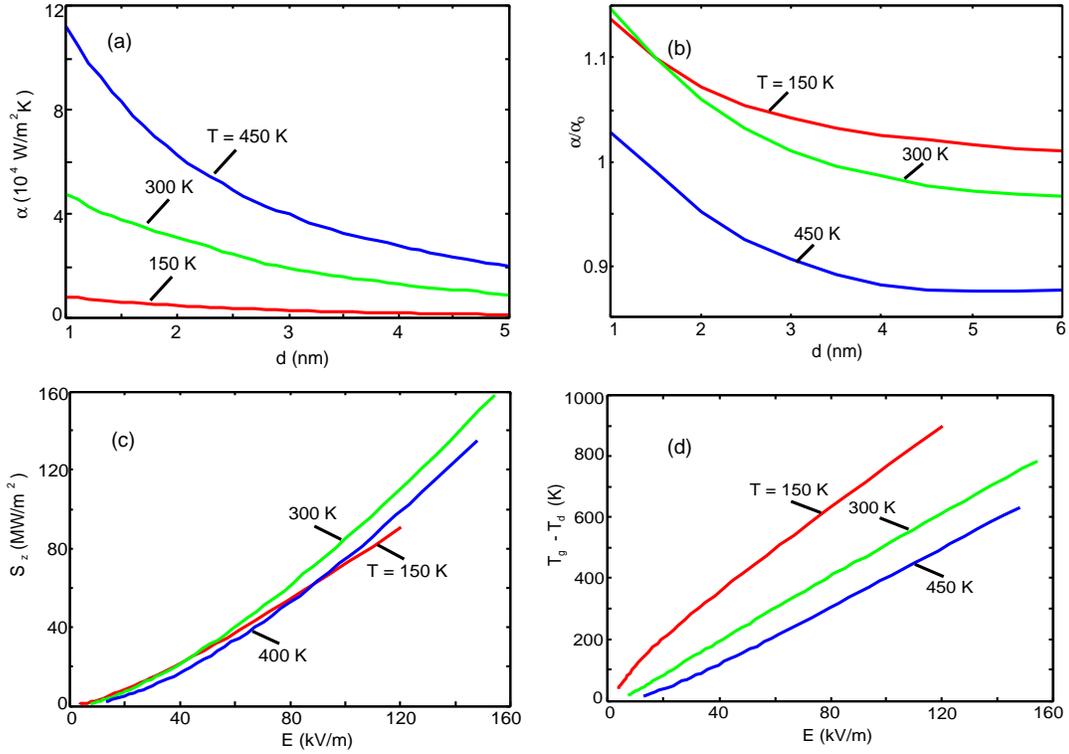}
\caption{\label{Fig.2.} Radiative energy transfer between graphene   and
SiO$_2$ for $n=10^{12}$cm$^{-2}$ and $\alpha_{ph}=0$. (a) Dependence of the
energy transfer coefficient on the separation $d$ for low electric
field ($v\rightarrow 0$).  (b) Dependence of the ratio between  the
energy transfer coefficient and the heat transfer coefficient on the separation $d$ for low electric
field ($v\rightarrow 0$). (c) Dependence of the radiative energy flux on electric field for $d=1.0$ nm. (d) Dependence of the temperature difference
between graphene and substrate on electric field for $d=1.0$ nm.}
\end{figure}

 Fig. \ref{Fig.2.}(a) shows the dependence of
the  energy transfer coefficient  on the separation $d$ for low
electric field ($v\rightarrow 0$). At $d \sim$ 5 nm and $T=300$ K the energy
transfer coefficient, due to the near-field radiative energy transfer, is
$\sim 10^4$Wm$^{-2}$K$^{-1}$, which is $\sim 3$ orders
of magnitude larger than the radiative heat transfer coefficient of the black-body radiation. In
comparison, the near-field radiative heat transfer coefficient  in
SiO$_{2}$-SiO$_{2}$ system for the plate-plate configuration, when
extracted from experimental data \cite{Chen2009a} for the
plate-sphere configuration, is $\sim $
2230Wm$^{-2}$K$^{-1}$ at a $\sim$30 nm gap. For this
system the radiative heat transfer coefficient depends on the separation as
$1/d^2$. Thus $\alpha \sim 10^5$Wm$^{-2}$K$^{-1}$ at
$d\sim 5$ nm, which is one order of magnitude larger than for the
graphene-SiO$_2$ system in the same configuration. However, the
sphere has a characteristic roughness of $\sim$ 40 nm, and the
experiments \cite{Chen2009a,Greffet2009} were restricted to
separation wider than 30 nm (at smaller separation the
imperfections affect the measured heat transfer). Thus the extreme
near-field-separation, with $d$ less than approximately 10 nm, may
not be accessible using a plate-sphere geometry. A suspended graphene sheet has a roughness $\sim$1 nm
\cite{Meyer2007}, and measurements of the thermal contact
conductance can be performed from separation larger than $\sim$ 1
nm. At such separation one would expect the emergence of nonlocal
and nonlinear effects. This range is of great interest for the
design of nanoscale devices, as modern nanostructures are
considerably smaller than 10 nm and are separated in some cases by
only a few Angstroms.

Fig. \ref{Fig.2.}(b) shows that at small separation there is significant difference
between the radiative energy transfer coefficient and the the radiative heat transfer coefficient determined (in absence of direct phononic coupling) by
$\alpha_0=S_{z0}^{\prime}$. This difference vanishes for large separation because $S_{z0}$ and $F_{x0}$ rapidly decrease when separation
increases. At large separation the friction force is dominated by intrinsic friction and in this case $\alpha\approx \alpha_0$. Fig. \ref{Fig.2.}(c)
shows the dependence of the radiative energy flux  on electric field for $d=1$ nm. For this separation the energy transfer is considerably less effective than for  $d=0.35$ nm, which leads to a rapid increase of the temperature difference (see Fig. \ref{Fig.2.}(d)). High temperatures are achieved at low electric field (small drift velocities) when contribution to the radiative energy transfer from quantum fluctuations is very small and the energy transfer is mainly determined by thermal fluctuations.

In conclusion, we have used theories of the van der Waals friction
and the near-field radiative energy transfer to study  heat
generation and dissipation in graphene due to the interaction with
phonon-polaritons in the (amorphous) SiO$_2$ substrate and acoustic phonons in graphene.
For the low-field energy transfer between nonsuspended graphene and
the substrate,  radiative energy transfer gives a significant contribution in addition to the phononic heat transfer. High-field heat
transfer is determined by the phononic mechanism. For high electric field (large drift velocities) and low temperatures
quantum fluctuations give an
important contribution to the energy flux and the friction force. For suspended
graphene the energy  transfer  coefficient at nanoscale gap is $\sim$ 3
orders of magnitude larger than the radiative heat transfer coefficient of the blackbody radiation
limit. We have pointed out that graphene can be used to study
near-field radiative heat transfer in the plate-plate
configuration, and for shorter separations than it is possible now
in the plate-sphere configuration.

 \vskip 0.5cm
\textbf{Acknowledgment}

A.I.V acknowledges financial support from the Russian Foundation
for Basic Research (Grant N 10-02-00297-à) and ESF within activity
``New Trends and Applications of the Casimir Effect''.

\end{document}